\documentclass[useAMS,usenatbib]{mn2e}

\usepackage{graphicx}

\title[First detailed analysis of multiple system V2083 Cyg]{First detailed analysis of multiple system V2083 Cyg}
\author[P.Zasche and P. Svoboda and M. \v{S}lechta]{P.Zasche$^{1}$\thanks{E-mail:
zasche@sirrah.troja.mff.cuni.cz} and P. Svoboda$^{2}$ and M. \v{S}lechta$^{3}$\\
$^{1}$Astronomical Institute, Faculty of Mathematics and Physics, Charles University Prague, CZ-180 00 Praha 8,
V Hole\v{s}ovi\v{c}k\'ach 2,\\ Czech Republic\\
$^{2}$Private observatory, V\'ypustky 5, Brno, CZ-614 00 Czech Republic\\
$^{3}$Astronomical Institute, Academy of sciences, Fri\v{c}ova 298, CZ-251 65, Ond\v{r}ejov, Czech
Republic}
\begin{document}

\date{Accepted ... Received ...; in original form ...}

\pagerange{\pageref{firstpage}--\pageref{lastpage}} \pubyear{2012}

\maketitle

\label{firstpage}

\begin{abstract}
Main aim of this paper is the first detailed analysis of multiple system V2083~Cyg and to reveal
its basic physical properties. The system was studied by method of the light and radial velocity
curves analysis, together with the interferometric data of the visual pair obtained during a last
century. There was found that the close subsystem contains two very similar stars of spectral type
A7-8. Moreover, the third body is orbiting around this pair with period of about 177~years. Due to
the discrepancy of total mass as derived from two methods, there arises that the third body is
maybe also a binary, or some object with lower luminosity but higher mass than normal main-sequence
star. Another explanation is that the Hipparcos value of parallax is incorrect and the system is
much closer to the Sun. 
\end{abstract}


\section{Introduction}

The eclipsing binaries as members of more complex multiple systems can provide us important
information about their physical properties, as derived from different methods. This is the case of
V2083~Cyg, which is the system, where the close components form an eclipsing binary, and the third
distant body orbiting the close pair is detected as a visual component. Thanks to the combined
analysis we are able to derive the radii, masses and evolutionary status of the close components
and also some properties of the distant one. Such systems are still very rare and mostly lie
relatively close to the Solar system. Nowadays, there are known only 33 such systems, where a close
eclipsing binary is a member of a wide visual binary and we know both orbits, their mutual
inclinations, ratio of periods, etc. Such a unique systems are the most suitable ones for studies
of dynamical effects, the short and long-term evolution of the orbits, etc. (see e.g.
\citealt{1975A&A....42..229S}).

\section{The system V2083~Cyg}

The system V2083~Cyg (= HD~184242 = HIP~96011, RA 19$^h$ 31$^m$ 16.36$^s$ DE~+47$^\circ$ 28$'$
52.24$''$, $V_{max} = 6.86$~mag) is an Algol-type eclipsing binary with its orbital period about
1.87~day. It is also a primary component of a visual double star designated as WDS J19313+4729 in
the Washington Double Star Catalog (WDS\footnote{http://ad.usno.navy.mil/wds/}, \citealt{WDS}). The
secondary component of this double star is about 220~mas distant and is a bit fainter. On the other
hand, the magnitude difference is not very certain, because different authors list different
values. The WDS catalogue itself gives 7.50 and 7.93~mag for both components.

The system is rather neglected one and there are only a few papers published. It was discovered as
an eclipsing binary from the Hipparcos data \citep{HIP}, which also reveal that the light curve
(hereafter LC) shows two similar minima and a classical feature of an Algol-type star.

Spectral type of the system is not known very precisely nowadays. \cite{1985ApJS...59...95A}
presented spectral classification of the whole AB system as Am (K/H/M=A3/A8/A9),
\cite{1991A&AS...89..429R} gives a composite spectral type as A3-A9, while the spectral type A3 was
presented by \cite{1918AnHar..91....1C}, \cite{1980BICDS..19...74O}, and many others. This could
indicate that the combined spectrum is composed from components of slightly different spectral
types. The photometry of V2083~Cyg obtained from the Hipparcos mission gives a color index
$B-V=0.279$~mag (indicating sp.type A9, \citealt{1980ARA&A..18..115P}), while the infrared $J-H$
and $H-K$ indices, which are less influenced by interstellar reddening, as derived from the 2MASS
survey give spectral types about A4 and A7, \cite{2000asqu.book.....C}.

The visual orbit of the two components was derived by \cite{Seymour2002V2083Cyg}. They presented
the orbital period of the double about 372~yr, the angular semimajor axis about 498~mas, and the
eccentricity 0.16. However, as they already mentioned, the orbit is still only a preliminary one.

\section{Photometry and spectroscopy}

We started collecting the photometric data of the system in April 2008. In total there are 31
nights of observations, but for the light-curve analysis we used only 27 nights of observations
obtained from April 2008 to September 2009 and carried out with the same telescope and detector at
the private observatory by one of the authors (PS). Owing to high brightness of the target, there
was used only a small 34-mm refractor at the Private observatory in Brno, Czech Republic, using the
SBIG ST-7XME CCD camera and standard $BVR$ filters by the specification by \cite{Bessell1990}. All
the measurements were processed by the software {\sc C-Munipack}\footnote{see
$\mathrm{http://c-munipack.sourceforge.net/}$}, which is based on aperture photometry and using the
standard DaoPhot routines \citep{Tody1993}. The other nights were used only for deriving the
precise times of minima for the system.

Besides our new observations, there were also used the photometric data obtained within the
SuperWASP survey \citep{2006PASP..118.1407P}. However, these data are not of enough quality to be
used for the LC analysis. Hence, we made use of the SWASP photometry only for deriving the minima
times of V2083~Cyg for a prospective period analysis. For all of the minima the Kwee - van Woerden
method was used \citep{Kwee}, and all of them are given in Table \ref{MINIMA}. The linear ephemeris
are as follows: $HJD = 2448501.1237 + 1.867493429 \cdot E.$

\begin{figure}
  \centering
  \includegraphics[width=80mm]{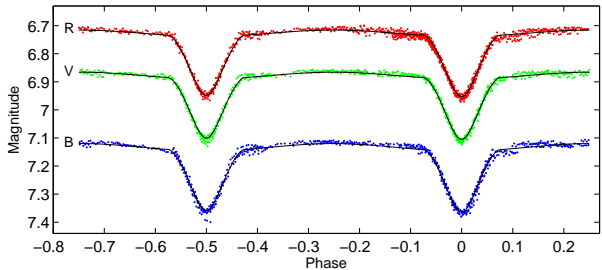}
  \caption{Light curves in $B$ (blue), $V$ (green), and $R$ (red) filters for V2083~Cyg, the solid
  line represents the final fit (see the text).}
  \label{Fig-LC}
\end{figure}

The CCD spectra were obtained at Ond\v{r}ejov observatory, Czech Republic, using the 2.0-m
telescope equipped with a SITe-005 800 $\times$ 2000 CCD detector. These spectra cover a wavelength
region 626 -- 676 nm. All of them were secured between April 2010 and May 2011 and have a linear
dispersion of about 17~$\AA$/mm. Their S/N values range typically between 100 and 200.

For all of the spectra used the wavelength calibration was made via a ThAr comparison spectra
obtained before and after the stellar spectra itself. The flatfields were taken in the beginning
and end of the night and their means were used for the reduction. After then, the radial velocities
(hereafter RV) were obtained with the program SPEFO \citep{1996A&A...309..521H}, with using the
zero point correction via measuring the telluric lines. In total 19 spectra were obtained this way.
Moreover, also two Elodie spectra \citep{2004PASP..116..693M} obtained in 1999 were added for the
analysis. 

A list of derived radial velocities from all of the available spectral observations is written in
Table \ref{RVs}. In the last column the reference Elodie or Ond\v{r}ejov is given. For all of the
spectra we also tried to identify the third component lines, however these radial velocities are
rather uncertain and affected by relatively large errors (see below Sections \ref{Visual} and
\ref{Physical}).

\section{LC and RV analysis}

The complete LC (in $BVR$ filters) and RV curves were analyzed simultaneously, using the program
{\sc PHOEBE} \citep{Prsa2005}, which is based on the Wilson-Devinney algorithm \citep{Wilson1971}.
The derived quantities are as follows: the semi-major axis $a$, the mass ratio $q = M_2/M_1$, the
systemic velocity $\gamma$, the secondary temperature $T_2$, the inclination $i$, the luminosities
$L_i$, the gravity darkening coefficients $g_i$, the limb darkening coefficients $x_i$, the albedo
coefficients $A_i$, and the synchronicity parameters $F_i$. The limb darkening was approximated via
linear cosine law, and the values of $x_i$ were interpolated from van~Hamme's tables, see
\cite{vanHamme1993}.

For the whole analysis, we followed this procedure: at the beginning, we fixed the temperature of
the primary component at $T_1=7930$~K (corresponding to spectral type A7,
\citealt{2000asqu.book.....C}). We were trying to find the best LC+RV fit according to the lowest
value of rms. There was a solution reached, but this one was unacceptable due to the fact that
resulting values of $M_1, M_2, L_1, L_2, T_1$, and $T_2$ are in contradiction with each other. In
particular, the resulting spectral types as derived from $M, L$ and $T$ differ significantly
between each other. For this reason, we tried a different starting value of $T_1$. With this method
we were changing the temperature $T_1$ in the range from 8520~K to 7020~K (spectral types A3 to F0)
and trying to find a consistent solution. For all of these attempts, the value of $T_1$ remained
fixed.

\begin{figure}
  \centering
  \includegraphics[width=80mm]{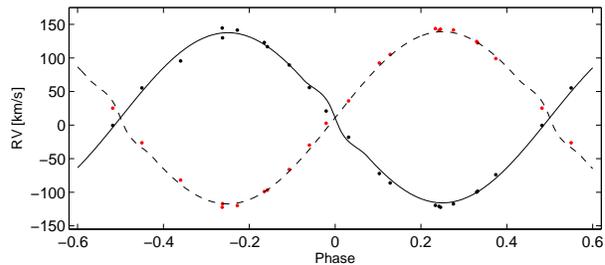}
  \caption{Radial velocity curves of V2083~Cyg for the primary (black points), and the secondary (red
  points), the lines represent the final fit (see the text and Table \ref{LCRVparam}).}
  \label{Fig-RV}
\end{figure}

Our final parameters as derived from the LC+RV fit are given in Table \ref{LCRVparam}. The plot of
the LC is shown in Fig. \ref{Fig-LC}, while the RV curves with the fits are given in Fig.
\ref{Fig-RV}. The value of eccentricity was fixed at 0. For the discussion about the physical
parameters of the components (eclipsing ones and also the third one), see Section \ref{Physical}.

For the whole computation process, the values of albedos $A_i$ and the values of gravity darkening
coefficients were set at their appropriate values ($A_i=1$ or 0.5, and $g_i=1$ or 0.32) according
to the component's temperatures ($T_i < 7200$~K, or $T_i > 7200$~K). Another problematic issue were
the values of $F_i$, which tended to decrease down to 0 for both components for each of the $T_1$
values. These dropped down very quickly after a few steps of iterations. Due to this reason we
tried a different approach. From the spectra of the system we estimated the values $v \sin i$,
which were used to derive the values of $F_i$ for both components. Therefore, the values of $F_i$
as given in Table \ref{LCRVparam} are not derived from the combined LC\&RV analysis but from the
spectra.

The fitting process with {\sc PHOEBE} was carried out assuming three luminosities. Besides the
luminosities of primary and secondary component of the eclipsing binary pair also the additional
third light $L_3$ was considered. This luminosity corresponds to the visual component B and is
presented in the combined light for all the time (the two visual components are too close). From
this value one can speculate about some physical parameters of the third body in the system, see
below Section \ref{Physical}.

  \begin{figure}
  \centering
  \includegraphics[width=80mm]{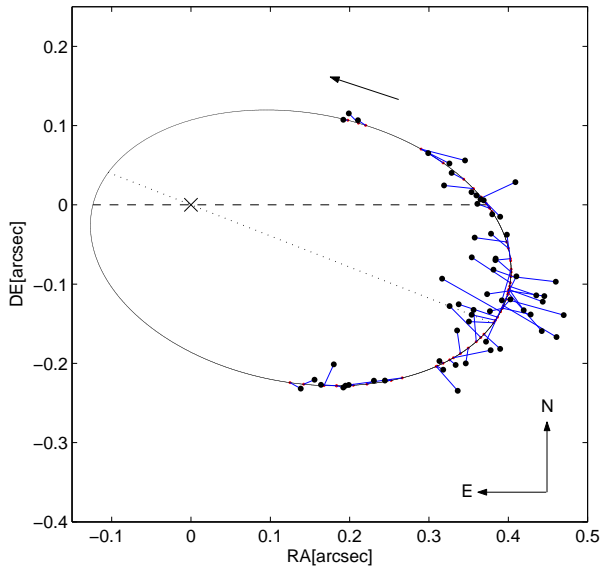}
  \caption{Visual orbit of V2083~Cyg as displayed on the sky. The individual observations
  (black points) are connected with their theoretical positions (red points). The dashed line
  represents the line of the nodes, while the dotted one the line of the apsides. Parameters of
  the fit are given in Table \ref{LongOrbit}.}
  \label{Fig-VisOrbit}
\end{figure}

\begin{table}
 \centering
 \begin{minipage}{80mm}
  \caption{Radial velocities of V2083~Cyg as derived from the spectra
  from the {\sc ELODIE} archive, and from the Ond\v{r}ejov observatory.}  \label{RVs}
  \begin{tabular}{@{}c c c c c@{}}
\hline
HJD - 2400000 & RV1 & RV2 & RV3 & Ref. \\
 \hline
 51405.3658 &  95.594 & -81.861 &         & Elodie \\
 51407.4156 & 129.872 & -117.06 &         & Elodie \\
 55316.5290 &  20.920 &   2.771 & -16.272 & OND \\
 55380.5210 &-122.394 & 142.771 & -16.336 & OND \\
 55385.3570 & 122.946 & -99.062 & -16.775 & OND \\
 55385.5530 &  56.001 & -29.860 & -18.597 & OND \\
 55386.3640 & -73.970 &  99.094 & -14.986 & OND \\
 55386.5650 &   -.442 &  25.184 & -17.640 & OND \\
 55405.3670 &  55.349 & -26.402 & -16.327 & OND \\
 55425.3970 &-117.384 & 141.598 & -16.145 & OND \\
 55496.2990 &-121.010 & 140.992 & -16.273 & OND \\
 55496.4630 & -99.677 & 124.485 & -16.277 & OND \\
 55497.2890 & 141.451 &-119.863 & -15.212 & OND \\
 55622.5420 & 116.769 & -96.782 & -15.048 & OND \\
 55622.6370 &  89.474 & -66.160 & -14.284 & OND \\
 55671.4500 & -18.008 &  36.024 & -14.574 & OND \\
 55671.5840 & -72.089 &  92.471 & -14.950 & OND \\
 55671.6310 & -86.135 & 105.183 & -14.784 & OND \\
 55689.5750 & 144.545 &-122.291 & -16.380 & OND \\
 55692.3700 &-119.493 & 143.646 & -15.231 & OND \\
 55692.5540 & -98.435 & 122.422 & -15.080 & OND \\
 \hline
\end{tabular}
\end{minipage}
\end{table}

\begin{table}
 \centering
 \begin{minipage}{80mm}
  \caption{The final LC and RV parameters of V2083~Cyg.}  \label{LCRVparam}
  \begin{tabular}{@{}c c c c@{}}
\hline
 Parameter & Value & Parameter & Value \\
 \hline
 $a$ [R$_\odot$] & 9.57 $\pm$ 0.15   & $L_1$ (B) [\%] & 26.7 $\pm$ 0.7 \\
 $q = M_2/M_1$   & 0.97 $\pm$ 0.07   & $L_2$ (B) [\%] & 34.9 $\pm$ 0.9 \\
 $\gamma$ [km/s] & 10.78 $\pm$ 0.68  & $L_3$ (B) [\%] & 38.4 $\pm$ 0.8 \\
 $T_1$ [K]       & 7630 (fixed)      & $L_1$ (V) [\%] & 26.7 $\pm$ 0.6 \\
 $T_2$ [K]       & 7623 $\pm$ 45     & $L_2$ (V) [\%] & 34.7 $\pm$ 0.9 \\
 $e$             &  0 (fixed)        & $L_3$ (V) [\%] & 38.6 $\pm$ 0.8 \\
 $i$  [deg]      & 80.47 $\pm$ 1.60  & $L_1$ (R) [\%] & 26.4 $\pm$ 0.6 \\
 $x_1=x_2$ (B)   & 0.412             & $L_2$ (R) [\%] & 34.2 $\pm$ 0.9 \\
 $x_1=x_2$ (V)   & 0.356             & $L_3$ (R) [\%] & 39.4 $\pm$ 1.0 \\    \cline{3-4}
 $x_1=x_2$ (R)   & 0.356             & \multicolumn{2}{c}{Derived physical quantities:} \\
 $g_1=g_2$       & 1.000 (fixed)     & $R_1$ [R$_\odot$] & 2.12 $\pm$ 0.17 \\
 $A_1=A_2$       & 1.000 (fixed)     & $R_2$ [R$_\odot$] & 2.45 $\pm$ 0.20 \\
 $F_1$           & 0.81 $\pm$ 0.13   & $M_1$ [M$_\odot$] & 1.71 $\pm$ 0.11 \\
 $F_2$           & 0.84 $\pm$ 0.11   & $M_2$ [M$_\odot$] & 1.66 $\pm$ 0.09 \\
 \hline
\end{tabular}
\end{minipage}
\end{table}

\section{Visual orbit} \label{Visual}

The close eclipsing pair is orbiting around a common barycenter with the third distant component of
the system. Recent precise interferometric observations are to be used for determining the
parameters of this visual orbit. Since its discovery as a double star by \cite{1904LicOB...3....6A}
there were obtained 61 astrometric observations of the double (i.e. position angle and separation).
We took these data from the WDS database. The very last observation was obtained in 2009.

Since its discovery, the position angle of the pair changed of about 88~$^\circ$. Thanks to this
movement, the orbit of the pair around a barycenter was derived. The orbit was published by
\cite{Seymour2002V2083Cyg}, who computed its orbital period of about 372~yr. However, since this
most recent study there were published three new interferometric observations, so we decided to
perform a new analysis with the complete data set.

\begin{figure}
  \centering
  \includegraphics[width=86mm]{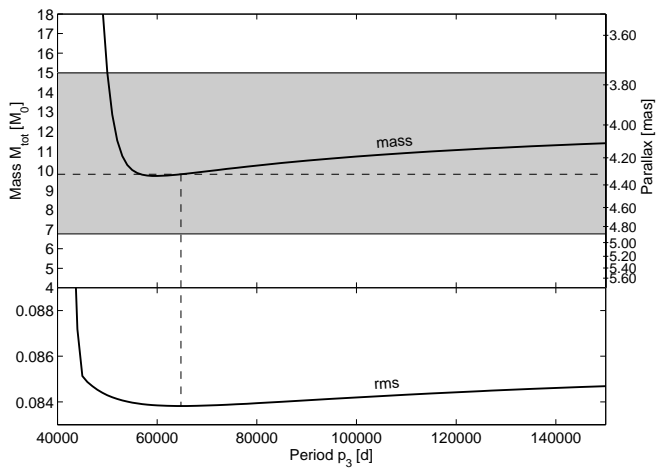}
  \caption{Plot of period-mass and period-rms relation. The dashed lines represent our final fit. For
  details see the text.}
  \label{chi2}
\end{figure}

Our new computation led to the visual orbit parameters given in Table \ref{LongOrbit}. For the
computation we used a following approach. Starting with the orbital parameters as published by
\cite{Seymour2002V2083Cyg}, the final fit reached very different solution. Moreover, there were
found several different minima in the parameter space as derived from this astrometric data set.
Some minimum was found with very long orbital periods, but this solution seems to be less probable
due to the poor coverage of the data. The most significant minimum (the deepest one) was found near
the period of 177.4~years. However, we would like to emphasize that the orbital solution is still a
preliminary one. New precise observations secured every year would be very welcome for derivation
of the orbital parameters with higher conclusiveness and especially for setting the more solid
constraints on $p_3$ and $a$ values. These are the most important for a discussion about the nature
of the third component (see Section \ref{Physical} below).

In Fig. \ref{chi2} there is a plot of total mass versus period, as well as the rms of the
particular fit versus period. For our final solution reached (minimum rms with $p_3 =
64778.357$~day) the value of total mass was computed (using the Hipparcos parallax) - these are
shown as dashed lines in Fig. \ref{chi2}. The relation between the two vertical axis (parallax and
total mass) is defined via a third Kepler's law using our final solution. As one can see from the
bold line of mass-period relation, the total mass as derived from our final solution is close to
the minimal mass in this period range (the uncertainty of Hipparcos parallax
$\pi_{HIP}=4.32\pm0.57$ is shown as a grey area). Of course, this analysis is very sensitive to the
input weightening scheme. The individual weights of the data points were set equal to each other,
because for most of the observations the sigma or some other error estimations are missing. No
minimum of rms near a period of 372~years as proposed by \cite{Seymour2002V2083Cyg} is presented.
One can ask, why such a different solution was reached when using only three new interferometric
observations. The main reason (maybe besides different weightening) is that these three new
measurements provide a strong constraints on the fit. This is due to the fact that the position
angle between our most recent data and those from \cite{Seymour2002V2083Cyg} changed about
20$^\circ$, which is about 1/4 of the total position angle range covered. All of these calculations
(e.g. the Kepler's law) were using the set of recommended values of fundamental parameters as
proposed by \cite{2011PASP..123..976H}.

On the other hand, we also tried to compute the predicted change of the third-body velocities over
the time span of more than 11 years covered with our spectroscopic data. Taken into account some
assumptions (masses), the change in velocity resulted in more than 20~km/s. Such a large velocity
difference should be easily detectable in our $RV_3$ data. Unfortunately, we were not able to
identify the third-component lines in the ELODIE spectra and in newer data from Ond\v{r}ejov there
is no such difference, hence we could only speculate about our findings. The reason could be either
different masses or much longer orbital period. Another explanation is an incorrect identification
of the third-body lines in the spectra.

\section{Physical parameters} \label{Physical}

Taken into account all results as presented above, one can make a picture of the system, its
geometry and orientation in space. From the combined LC and RV analysis there resulted that both
eclipsing components are probably main sequence stars, located well within their respective Roche
lobes. According to their masses and temperatures, it seems like their individual spectral types
are probably of A7 and A8 (e.g. \citealt{1980ARA&A..18..115P}, \citealt{Hec1988}, or
\citealt{1991A&ARv...3...91A}) for primary and secondary, respectively. However, according to their
luminosities, it seems like the stars are of slightly earlier spectral type (about A5).

Another task was to derive the value of the third light $L_3$ from the LC solution and to obtain a
magnitude difference between the two visual components. This value resulted in about 0.49~mag,
which is in rough agreement with the $\Delta m = 0.43~$mag value as presented in the WDS catalogue.

\begin{table}
 \centering
 \begin{minipage}{80mm}
  \caption{Final parameters of the long orbit.}  \label{LongOrbit}
  \begin{tabular}{@{}c c c c@{}}
\hline
 Parameter  & \cite{Seymour2002V2083Cyg} & This work \\
 \hline
 $p_3$ [day]    & 135869  & 64778 $\pm$ 427    \\
 $p_3$ [yr]     & 372     & 177.4 $\pm$ 1.2    \\
 $a$  [mas]     & 498     & 291.9 $\pm$ 1.4    \\
 $T_0$          & 2438395 & 2400006 $\pm$ 375  \\
 $\Omega$ [deg] & 73.6    & 174.54 $\pm$ 2.9   \\
 $\omega$ [deg] & 189     & 334.89 $\pm$ 5.3   \\
 $i$ [deg]      & 64      & 48.73 $\pm$ 3.6    \\
 $e$            & 0.16    & 0.471 $\pm$ 0.018  \\
 \hline
\end{tabular}
\end{minipage}
\end{table}

A discussion about the third body is still difficult due to some aspects of the problem. The most
problematic issue is still the uncertainty of the Hipparcos value of parallax. The relatively high
error of about 13~\% could lead to distances in wide range from 204 to 267~pc. Thanks to this
uncertainty also the value of total mass as computed from the visual orbit (see Table
\ref{LongOrbit}) lie in between 6.54 and 15.41~M$_\odot$ with the mean value of 9.81~M$_\odot$.

Subtracting masses of both eclipsing components, we obtain an interesting result of mass of the
third body about 6.44~M$_\odot$ (with upper and lower limits about 12.24 and 2.97). Such a massive
third body easily cannot be a main sequence A star as predicted from the $\Delta m$ value. One
possible explanation of this discrepancy is that this component is also a double star. If we
speculate about two identical stars, then such stars have to be of only slightly later spectral
type than the eclipsing components. (because of the total mass). Assuming two F0 stars, we can
hardly satisfy the magnitude difference between the components. However, this explanation is still
questionable because the third lines in the spectra do not show a doubleness profile.

To solve this discrepancy we also tried to use the program {\sc KOREL} \citep{2004PAICz..92...15H}
for disentangling of the spectra taken at Ond\v{r}ejov observatory. However, it was also not able
to solve the problem. The final parameters on one hand confirmed our findings about the LC+RV
solution (mass ratio $q$ from {\sc KOREL} resulted in about 0.993), but on the other hand it also
results with the value of mass ratio $q_3 = M_3/M_{12} > 1$. This would indicate that the third
body is more massive than the eclipsing pair, but also the less luminous one. Solving the problem
of its lower luminosity and higher mass with introducing a degenerate object is a highly
speculative solution. Hence, the nature of the third body still remains an open question. The {\sc
KOREL} radial velocities of the third body were also used and these are the values presented in
Table \ref{RVs} in the $RV3$ column.

\section{Discussion and conclusions}

The multiple system V2083~Cyg is still rather neglected one and this is the first detailed analysis
of it. The components of an eclipsing binary are of spectral type A, are well-detached, with no
evidence of circumstellar matter, emission in the spectra, etc. This close pair is also orbiting
around a common barycenter with the third component with period about 177~years. The mutual
inclination of the two orbits is 31.8$^\circ$, therefore we can only speculate about a common
origin of the system.

A nature of the third component is still rather problematic to derive. From the combined LC and RV
analysis there results that the third body is slightly less luminous than the eclipsing pair. But
the Hipparcos parallax indicates a higher total mass of the system than computed from all
components' masses. A possible explanation is that the value of Hipparcos parallax is
underestimated and a real distance of V2083~Cyg is lower (even outside of the error bars of the
Hipparcos data). This would not be an exceptional case, because for some systems the Hipparcos data
yielded incorrect parallax due to the presence of a close visual companion (e.g.
\citealt{2008AJ....136..890D}). Another possible explanation is that the body is also a binary, but
there are also some problems with this explanation (luminosity and the spectral lines of such
body). For that reason, new more detailed observations are still very welcome.

However, if the hypothesis of binarity of the third component is proved, it will shift the triple
to quadruple. On one hand, such systems of higher multiplicity are of big interest, on the other
hand we also deal with very incomplete statistics of them among the stars (see e.g.
\citealt{2008MNRAS.389..869E}, and \citealt{2009MNRAS.399.1471E}).

\section*{Acknowledgments}

We thank the "SWASP" team for making all of the observations easily public available. Mr. Robert
Uhla\v{r} is acknowledged for sending us his photometric data and Prof. Petr Harmanec for a useful
discussion and valuable advices. We also thank an anonymous referee for his/her helpful and
critical suggestions. We also acknowledge the use of spectrograms from the public archives of the
Elodie spectrograph of the Haute Provence Observatory. We are obliged to our colleagues who took
spectra of V2083~Cyg in Ond\v{r}ejov observatory (M. Wolf, P. Chadima, and J.A. Nemravov\'a). This
work was supported by the Czech Science Foundation grant no. P209/10/0715 and also by the Research
Programme MSM0021620860 of the Czech Ministry of Education. This research has made use of the
Washington Double Star Catalog maintained at the U.S. Naval Observatory, the SIMBAD database,
operated at CDS, Strasbourg, France, and of NASA's Astrophysics Data System Bibliographic Services.

\appendix

\section[]{}

\label{lastpage}

\begin{table}
 \centering
 \small
 \begin{minipage}{80mm}
  \caption{Heliocentric minima of V2083~Cyg.} \label{MINIMA}
  \begin{tabular}{@{}l l l l l l@{}}
\hline
HJD - 2400000 & Error & Type & Filter & Observer\\
 \hline
 54609.69354 & 0.00179 & prim & - & SWASP \\
 54638.63794 & 0.00035 & sec  & - & SWASP \\
 54639.57227 & 0.00042 & prim & - & SWASP \\
 54652.64554 & 0.00089 & prim & - & SWASP \\
 54668.51720 & 0.00046 & sec  & - & SWASP \\
 54669.44747 & 0.00046 & prim & - & SWASP \\
 54683.46156 & 0.00027 & sec  & - & SWASP \\
 54684.39467 & 0.00095 & prim & - & SWASP \\
 54994.39887 & 0.00091 & prim & B & PS \\
 54994.39888 & 0.00113 & prim & V & PS \\
 54994.39948 & 0.00128 & prim & R & PS \\
 55049.49292 & 0.00110 & sec  & B & PS \\
 55049.49015 & 0.00104 & sec  & R & PS \\
 55049.49055 & 0.00062 & sec  & V & PS \\
 55051.35585 & 0.00081 & sec  & B & PS \\
 55051.35567 & 0.00078 & sec  & V & PS \\
 55051.35645 & 0.00107 & sec  & R & PS \\
 55064.43157 & 0.00091 & sec  & B & PS \\
 55064.43162 & 0.00057 & sec  & V & PS \\
 55064.43102 & 0.00064 & sec  & R & PS \\
 55076.57570 & 0.00116 & prim & B & PS \\
 55076.57197 & 0.00108 & prim & V & PS \\
 55076.57194 & 0.00165 & prim & R & PS \\
 55093.37749 & 0.00087 & prim & B & PS \\
 55093.37848 & 0.00068 & prim & V & PS \\
 55093.37624 & 0.00063 & prim & R & PS \\
 55374.43431 & 0.00021 & sec  & I & RU \\
 55429.52450 & 0.00028 & prim & I & RU \\
 55740.46289 & 0.00056 & sec  & R & PS \\
 55797.42284 & 0.00043 & prim & I & RU \\
 \hline
\end{tabular}
 \begin{list}{}{}
 \item[] Note: PS - Petr Svoboda, RU - Robert Uhla\v{r}
 \end{list}
\end{minipage}
\end{table}

\end{document}